\documentclass[11pt]{gh2013}

\usepackage{mathptmx}       
\usepackage{helvet}         
\usepackage{courier}        
\usepackage{makeidx}         
\usepackage{graphicx}        
\usepackage{multicol}        

\begin{document}

\def\Ha{H$\alpha$}
\def\hii{H{\sc ii}}
\def\hi{H{\sc i}}
\def\oiii{[O{\sc iii}]}
\def\sii{[S{\sc ii}]}
\def\siii{[S{\sc iii}]}
\def\ergs{{\rm\,erg\,s^{-1}}}
\def\fesc{$f_{\rm esc}$}
\def\Lha{$L_{\rm H\alpha}$}
\def\nhi{$N$(\hi)}

\title*{Ionization by Massive Young Clusters as Revealed by Ionization-Parameter Mapping}
\titlerunning{\hii\ Region and Starburst Optical Depths}

\author{M. S. Oey$^1$,
E. W. Pellegrini$^2$,
J. Zastrow$^1$, and
A. E. Jaskot$^1$}
\authorrunning{Oey, Pellegrini, Zastrow, \& Jaskot}
\institute{$^{1}$Department of Astronomy, University of Michigan,
Ann Arbor, MI\ \ \ 48109-1042, USA
\\ $^{2}$Department of Physics and Astronomy, University of Toledo,
Toledo, OH, 43606, USA}
%
%
\maketitle


\vskip -3.5 cm  
\abstract{
Ionization-parameter mapping (IPM) is a powerful technique for tracing
the optical depth of Lyman continuum radiation from massive
stars. Using narrow-band line-ratio maps, we examine trends in
radiative feedback from ordinary \hii\ regions of the Magellanic Clouds
and nearby starburst galaxies. We find that the aggregate escape
fraction for the Lyman continuum is sufficient to ionize the diffuse,
warm ionized medium in the Magellanic Clouds, and that more luminous
nebulae are more likely to be optically thin.  We apply
ionization-parameter mapping to entire starburst galaxies, revealing
ionization cones in two nearby starbursts.  Within the limits of our
small sample, we examine the conditions for the propagation of
ionizing radiation beyond the host galaxies. 
}

\section{Introduction}

Massive young clusters are responsible for giant regions of ionized
gas.  These are photoionized by the clusters themselves, as
well as shock-ionized by supernovae and supersonic stellar
winds from the high-mass stars.  Here, I will focus on the
photoionized gas, which provides the emission-line diagnostics that
are widely used to evaluate a variety of phenomena across the
universe, for example, star-formation
rate and ionizing stellar populations, as well as
gas properties such as metallicity, density, pressure,
and kinematics.  Many papers presented in this conference exploit this
technique, in particular, work by E. Telles and M. Rodriguez in
this session.  When considering global properties of
galaxies, the \hii\ region luminosity function provides a quantitative
parameterization of star formation.  And finally, 
stellar radiation from optically thin \hii\ regions is responsible
for ionizing the interstellar medium and thereby generating the
diffuse, warm ionized medium (WIM).

It is therefore apparent that photoionization by massive stars in
clusters not only provides essential diagnostics of physical
conditions, but also is itself a fundamental process.
What is the fate of ionizing photons?  We clearly see \hii\ regions in
the immediate vicinity of massive clusters, but a significant
proportion of these regions must be optically thin, and thus
ionize the WIM.  Similarly, if the ISM is itself optically thin in 
some galaxies, then Lyman continuum radiation will escape into the
circumgalactic medium and perhaps, the intergalactic medium.  It is
widely believed that this was the case in early cosmic times, and that
massive stars are responsible for the reionization of the universe.
Thus, photon path lengths likely range across many orders of
magnitude.  It is also important to keep in mind that when the photons
are absorbed, whether near or far from their source, they not only
cause ionization of matter, but also impart momentum, generating
radiation pressure that may be significant to gas kinematics.  This
topic is explored by S. Silich in this session.

\section{Ionization-Parameter Mapping}

A central issue is therefore quantifying which \hii\ regions are optically
thin.  Many studies have examined this problem using different
strategies.  While some authors model globally-integrated
emission-lines (e.g., Iglesias-P\'aramo \&
Mu\~noz-Tu\~non 2002; Giammanco et al. 2004), others
model the WIM surface brightness generated by
its opacity and scattering (e.g., Zurita et al. 2002; Seon 2009).
Here, we demonstrate an elegant technique to evaluate the optical depth of
\hii\ regions based on ionization-parameter mapping (IPM),
exploiting the fact that the nebular ionization structure varies
between optically thin and optically thick objects.

\begin{figure*}
  \centering
  \includegraphics[width=  5 cm]{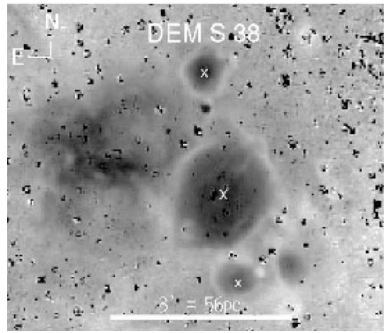}
  \caption{Ratio map for \oiii$\lambda5007$/\sii$\lambda6717$ of
    SMC nebulae; black and white indicate high and low values,
    respectively.  The objects marked by crosses exhibit classic,
    Str\"omgren sphere ionization structure.  (N is up, E to the left;
    from Pellegrini et al. 2012.)
}
  \label{f_neb}
\end{figure*}

Figure~\ref{f_neb} shows the \oiii/\sii\ emission-line ratio map for some
\hii\ regions in the Small Magellanic Cloud (SMC).  In the case of
an optically thick, Str\"omgren sphere, the ionization state will be high in
the central regions near the ionizing source, and will decrease at
large radii, as it transitions to the neutral environment at the
nebular boundary.  Thus, the nebular structure shows more
highly ionized species, represented in Figure~\ref{f_neb} by O$^{++}$, in the
center, surrounded by an envelope of lower-ionization species like
S$^+$.  The round \hii\ regions marked with crosses in Figure~\ref{f_neb} demonstrate this
classic, Str\"omgren sphere ionization structure.  In contrast, the
object with irregular morphology to the east in Figure~\ref{f_neb} appears to be
completely dominated by \oiii\ and is missing the low-ionization
envelope.  This implies that the object is density-bounded and
therefore optically thin.  The irregular morphology is also consistent
with radiation hydrodynamic simulations by Arthur et al. (2011)
showing that objects with the
highest ionization parameters, which are the ones that become
optically thin, develop instabilities leading to irregular
morphology such as that seen in the eastern object of Figure~\ref{f_neb}.

However, exploration of the parameter space with simple photoionization
models shows that this elegant picture is a bit more complicated.  
Using CLOUDY (Ferland et al. 1998)
models, we demonstrated that for lower ionizing effective temperatures,
low-ionization envelopes may be present even for quite optically thin
conditions (Pellegrini et al. 2012; hereafter P12).  Thus, for IPM
based on only two ionic species, it is the 
{\it absence} of these low-ionization transition zones that implies low
optical depth; whereas their presence merely implies a substantial likelihood
of optically thick conditions.  However, we note that optically thin
objects that possess low-ionization envelopes occur for cooler stars,
which tend to yield smaller and fainter \hii\ regions.  And as noted
above, nebular morphology is also a diagnostic criterion.
We tested IPM as an estimator of optical depth by applying the
method to a dozen nebulae in the Large Magellanic Cloud (LMC) with known
spectral classifications of the ionizing stars.  We crudely estimated
escape fractions \fesc\ by assigning the objects as optically thick,
thin, or blister objects based on their ionization structure.  These
categories were simply assigned values of \fesc\ $= 0$, 0.6, and 0.3,
respectively.  More accurate \fesc\ for these objects were then measured by
comparing the predicted versus observed \Ha\ luminosities.  The
extremely crude IPM estimates agree surprisingly
well with the measured values, to 25\% on average (P12).

So in general, we suggest that objects that {\it look} like Str\"omgren
spheres usually {\it are} Str\"omgren spheres.  Note that any
low-ionization envelopes should also be detected in the line of sight;
thus IPM diagnostics are not limited to projected radial analysis.
Furthermore, we stress 
that with three or more radially varying ionic species, the resulting
maps of the ionization structure will constrain the optical depth far
more strongly, essentially allowing its direct measurement.  We have
obtained such data for M33, which will permit us to quantify this
capability.

\section{Optical Depth of \hii\ Regions in the Magellanic Clouds}

The data presented above for individual nebulae in the LMC and SMC
were generated from the Magellanic Clouds Emission-Line Survey (MCELS;
Smith et al. 2005), a narrow-band imaging survey in \Ha, \oiii, and
\sii\ of the entire star-forming extent of both Clouds.  These data
allow us to apply IPM to crudely estimate the optical depths of the entire
\hii\ region population in both galaxies.  It is especially
interesting to quantitatively compare these, given that the LMC and SMC
have strongly contrasting neutral \hi\ ISM:  as seen in the
\hi\ surveys of these galaxies (Kim et al. 1998; Stanimirovi\'{c} et
al. 1999), the LMC has a highly shredded, filamentary neutral ISM
with high porosity, owing to extensive mechanical feedback
from massive stars and its flat, disk structure; whereas the SMC has a
more three-dimensional and diffuse \hi\ distribution with much lower
porosity due to its much lower specific star-formation rate (Oey
2007).  

\begin{figure*}
  \centering
  \includegraphics[width=3.5 cm]{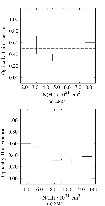}
\hspace*{2.0 cm}
  \includegraphics[width=3.5 cm]{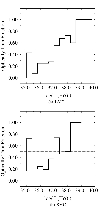}
  \caption{[Left] Frequency of optically thin objects as a function of
    \hi\ column density observed in the \hii\ region aperture, for the
    LMC (top) and SMC (bottom).  (From P12.)
}
  \label{f_tauHI}
\end{figure*}

Using IPM, we crudely categorize all \hii\ regions as optically
thick, thin, or neither (see P12
for details).  Figure~\ref{f_tauHI} shows the frequency of optically
thin objects as a function of \hi\ column density \nhi\ measured
within the nebular apertures for the LMC and SMC.  As expected, 
the fraction of optically thin objects clearly decreases with \hi\ column.
However, it is noteworthy that that the optically thin objects
actually dominate at the lowest \nhi\ in both galaxies.  The value of
\nhi\ at which this occurs is higher in the SMC than in the LMC, owing
to the SMC's 3-D ISM structure.  Furthermore,
we also see that there are still optically thin objects even at the
highest \hi\ columns.  While the mean \nhi\ for the optically thick
objects is larger in both galaxies, the \nhi\ distributions for the
optically thick and thin objects are similar in shape (P12). 

\begin{figure*}
  \centering
\vspace*{-1.2 cm}
  \caption{[Right] Frequency of optically thin objects as a function of
    \Ha\ luminosity, for the LMC (top) and SMC (bottom).  (From P12.)
}
  \label{f_tauHa}
\end{figure*}

Figure~\ref{f_tauHa} shows the frequency of optically thin objects as
a function of \Ha\ luminosity \Lha\ for the LMC and SMC.
In both galaxies, there is a clear trend that the
frequency increases with \Lha.  Even so, we again note
optically thick objects exist even at the highest
values of \Lha.  However, optically thin objects dominate in numbers
above a value of about $\log$ \Lha$\sim 37/\ergs$ in both galaxies.
This low luminosity corresponds to objects ionized by single,
mid-type O stars, implying that most of the bright \hii\ regions seen
in star-forming galaxies tend to be optically thin.  The
\Lha\ distributions for the optically thick vs thin objects differ far
more strongly than the distributions of their \nhi\ (P12).

The statistics for these populations indicate lower
limits on the total nebular escape fraction of \fesc $=0.42$
and 0.40 in the LMC and SMC, respectively.  From these, we can derive
total ``escape luminosities'' of $\log L_{\rm esc}/\ergs \sim 40.1$ and 39.2,
respectively, representing the total potential \Ha\ luminosities
allowed by these \fesc.  Comparing $L_{\rm esc}$ to observed WIM
luminosities in the LMC and SMC of $\log$ \Lha/$\ergs=40.0$ and 39.3,
respectively, we find that not only are the $L_{\rm esc}$ large
enough to fully ionize the WIM, but also that the global escape fractions
from these galaxies may be non-zero, when accounting for contributions
from field OB stars having no associated nebulae (see
P12 for details).  If this is the case, IPM shows
that \fesc\ is likely dominated by a few objects in the
galactic periphery, rather than the most luminous objects in the
dominant, central star-forming regions.

\section{Starburst Galaxies}

The individual Magellanic Clouds nebulae show that the
most luminous objects are more likely to be optically thin.  If
so, then entire starburst galaxies might
also plausibly have large \fesc.  However, many studies have
evaluated this possibility with mixed results.  Only two local
starbursts have confirmed detections of the Lyman continuum (e.g.,
Leitet et al. 2013; Grimes et al. 2009), while a minority of
Lyman-break galaxies show such detections (e.g., Iwata et al. 2009;
Shapley et al 2006).  Absorption-line studies of local starbursts all
imply optically thick conditions (e.g., Heckman et al. 2001; Leitherer
et al. 1995).

\begin{figure*}
  \centering
  \includegraphics[width= 5.5 cm]{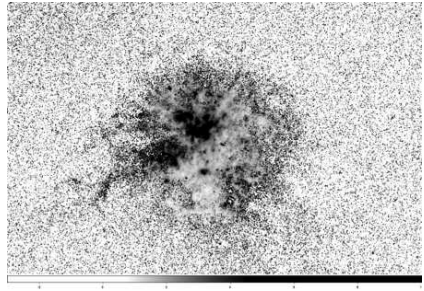}
  \caption{Ratio map for \siii$\lambda9069$/\sii$\lambda6717$ of the
    starburst galaxy Haro~10 (NGC 5253).  Black and white correspond
    to high and low values, respectively.  An ionization cone is
    apparent, oriented toward the southeast.  (N is up and E to the
    left; from Zastrow et al. 2011.)
}
  \label{f_Haro10}
\end{figure*}

We therefore apply the IPM technique globally to a sample of local starburst
galaxies:  Haro 10 (NGC 5253), NGC 3125, Henize 2-10, NGC 1705, and NGC
178 (Zastrow et al. 2013).  We carry out mapping in \siii\
and \sii, using the Maryland-Magellan Tunable Filter (MMTF) at Magellan.
The IPM technique is again proven, revealing a vivid ionization cone in
Haro 10 (Zastrow et al. 2011; Figure~\ref{f_Haro10}) and NGC 3125.
The remaining galaxies show no significant evidence of optically thin regions.
The narrow morphology of the ionization cones suggests that galaxy
orientation to the line of sight plays a major role in the detection
of Lyman continuum and optically thin gas.  Thus, significant values
of galactic \fesc\ may be more common than observations thus far
imply.

While our sample is small, our results are also consistent with
suggestions that a mininum star-formation intensity is needed to
generate optically thin conditions (e.g., Fernandez \& Shull 2011).
It is also likely that a specific age range is necessary, 
corresponding to times after which mechanical feedback from the
starburst has sufficiently shredded the ISM to facilitate the escape of
ionizing radiation, but also before the ionizing stars have expired.
This should be around 3 -- 5 Myr, again consistent with the data from
our small sample.  

Finally, since IPM implies that optically thin objects generate
enhanced ionization parameters, we suggest that objects with extreme
ionization parameters may therefore have high \fesc.  The
presentation in a later session by A. Jaskot demonstrates that this
scenario may indeed be likely (Jaskot \& Oey 2013).  Hence we see that
the technique of ionization-parameter mapping is a powerful tool on
both local and galaxy-wide scales.

\begin{acknowledgement}
Many thanks to the conference organizers for the opportunity to present this
work, which was supported by NSF AST-1210285.

\end{acknowledgement}

\end{document}